\documentclass[3p,preprint]{elsarticle}
\usepackage{etoolbox}
\usepackage{amsmath,amssymb}
\usepackage{lineno,hyperref,mathtools, subfigure}
\modulolinenumbers[0]
\usepackage{subfigure}
\journal{Physica B: Condensed Matter}
\usepackage{bm}
\usepackage{multirow}
\usepackage{color}
\usepackage{tabularx}
\DeclareUnicodeCharacter{2212}{-}
\DeclareUnicodeCharacter{2009}{-}
\DeclareUnicodeCharacter{03BA}{-}
\begin{document}

\begin{frontmatter}

\title{Theoretical Study of Nonlinear Absorption of a Strong Electromagnetic Wave in Infinite Semi-parabolic plus Semi-inverse Squared Quantum Wells by Using Quantum Kinetic Equation}
\author[1]{Cao Thi Vi Ba}
\author[1]{Nguyen Quang Bau\corref{cor1}}
\author[1]{Anh-Tuan Tran}
\author[1]{Tang Thi Dien}

\cortext[cor1]{nguyenquangbau54@gmail.com, nguyenquangbau@hus.edu.vn}
\address[1]{Department of Theoretical Physics, University of Science, Vietnam National University, Hanoi, Address: $\rm{No}$ 334 Nguyen Trai, Thanh Xuan, Hanoi, Vietnam}

\begin{abstract}
General analytic expressions for the total absorption coefficient of
strong electromagnetic waves caused by confined electrons in Infinite semi-parabolic plus Semi-inverse Squared Quantum Wells (ISPSISQW) are obtained by using the quantum kinetic equation for electrons in the case of electron-optical phonon scattering. A second-order multi-photon process is included in the result. The dependence of the total absorption coefficient on the intensity $E_0$, the photon energy $\hbar \Omega$ of an SEMW, and the temperature T for a specific GaAs/GaAsAl ISPPSISQW is achieved by using a numerical method. The computational results demonstrate that the total absorption coefficient's dependence on photon energy can be utilized for optically detecting the electric sub-bands in an ISPPSISQW. Besides, we also give theoretical rules on the dependence of the Full Width at Half Maximum on important external parameters such as temperature and perpendicular magnetic field. Furthermore, the obtained results are consistent with prior theoretical and experimental findings.
\end{abstract}
\begin{keyword}
Total absorption coefficient \sep Infinite semi-
parabolic Plus Semi-inverse Squared Quantum Wells \sep Multiphoton absorption processes \sep Quantum kinetic equation \sep Strong Electromagnetic Wave \sep Full Width at Half Maximum
\end{keyword}

\end{frontmatter}

\section{Introduction}
Since the 1970s, Zhores I. Alferov and Herbert Kroemer developed heterostructured semiconductor systems to improve the speed of information transmission in photonics. In the beginning of the 21st century, Zhores I. Alferov and Herbert Kroemer were awarded the Nobel Prize in Physics for their important contributions to heterostructured semiconductor systems \cite{nobel1,nobel2}.

In recent years, solid-state physicists have continued to study the optoelectronic properties of heterostructures formed from GaAs and GaAsAl semiconductor layers both theoretically and experimentally. Among them, special attention is paid to two-dimensional electron-gas systems, a model made up of a nm-thin layer. This thin layer consists of a semiconductor with a small band gap, sandwiched between layers of other semiconductors with larger band gaps. The difference between the conduction band minima between the two semiconductors forms a confinement quantum well, restricting the movement of electrons in a certain direction (usually the z-direction). Thus, electrons can only move freely in the x - y plane, so the electronic system can be viewed as a real two-dimensional system. Depending on the different confinement potentials, the electron's behavior will be different and the physical properties exhibited by the electron system will also be different. Heterostructures with such quantum confinement potentials are called low-dimensional semiconductor systems. Theoretical studies on low-dimensional semiconductor systems have been of interest to physicists since the 70s of the 20th century demonstrated through many studies on the optical and electrical properties of the system under the influence of the external field, especially problems involving low-dimensional systems placed under the influence of a strong external electromagnetic wave \cite{smcdt1,smcdt2,smcdt3,smcdt4,dien}. 
At present, there are many studies on low-dimensional semiconductor systems with different confined potential profiles. For instance, previous studies on spherical quantum dots \cite{qd1,qd2}, quantum rings \cite{qr1,qr2}, elliptical or cylindrical quantum wires \cite{qw1,qw2,qw3}, and double triangular and symmetric double semi-parabolic quantum wells \cite{qtw1,qtw2} have all shown the significant influence of geometric structure on the nonlinear optical absorption effect. Indeed, to this day, the optical properties of low-dimensional semiconductor systems continue to attract widespread research interest among physicists, offering many potential new breakthrough applications in technology.

Numerous theoretical and experimental studies of linear and nonlinear optical absorption coefficients in low-dimensional systems have been conducted in the past, in which, theoretical calculations are performed mainly using Boltzmann classical kinetic equations \cite{eps, eps1}. Based on the framework of perturbed theory, Huynh Vinh Phuc and co-workers \cite{phuc2015,tung2} found the explicit expression of optical absorption power in Parabolic Quantum Well and Asymmetric Finite Quantum Well in the assumption of non-degenerate electron gas under a perpendicular magnetic field and proved that Phonon-assisted cyclotron resonance-linewidth varies with the square root of temperature. In previous works \cite{bau, bau1, bau4}, we also investigated the problem of nonlinear optical absorption in two-dimensional semiconductor systems, especially quantum wells with different potential profiles by using Quantum Kinetic Equation. However, the problem of absorbing strong electromagnetic waves in Quantum Wells with Infinite semi-parabolic plus Semi-inverse Squared confinement potential has not yet been studied. 

An important effect in investigating optical absorption problems of low-dimensional semiconductors is the electron-phonon resonance (EPR) effect. This effect is widely studied both theoretically and experimentally in bulk semiconductors \cite{eps1}, quantum wells \cite{bau2, bau4, lvtung}, superlattices \cite{phong} and graphene \cite{tuan2023}. The principal peak occurs in this effect when the photon energy of the electromagnetic wave is equal to the optical phonon energy $\hbar \Omega = \hbar \omega_{0}$ \cite{bau}. In previous studies, mainly the electron-phonon resonance effect was studied for the absorption of a photon in the confinement potentials in the form of squares, triangles, parabolas, ... without taking into account multi-photon absorption. Multi-photon absorption occurs when a material is capable of absorbing two or more photons simultaneously. This is a very important absorption process being studied by many scientists in recent years \cite{tang, kawata}. To contribute to the completion of multi-photon absorption studies related to the important quantum kinetic effects of electrons in low-dimensional semiconductor systems under the influence of different types of confinement potentials, we study a two-dimensional quantum well model where the electron is confined in an Infinite Semi-parabolic Plus Semi-inverse Squared Quantum Wells (ISPPSISQW).  

We present the rest of the paper as follows. In Section 2, we give the Quantum Well model along with the wave function and the discrete energy spectrum of the electron, and then we set up and solve the quantum kinetic equation, thereby proposing the analytic expression of a multiphoton total absorption coefficient (TAC) of strong electromagnetic waves (SEMW) in both cases with and without the magnetic field. In Section 3, based on the obtained analytic expression, we perform numerical calculations to plot graphs and give physics discussions to elucidate important properties. The conclusion is given in section 4.
\section{Theory}

\subsection{The wave function and the discrete energy spectrum of the electron in ISPPSISQW}
\begin{figure}[!htb]
    \centering
    \includegraphics[scale=0.7]{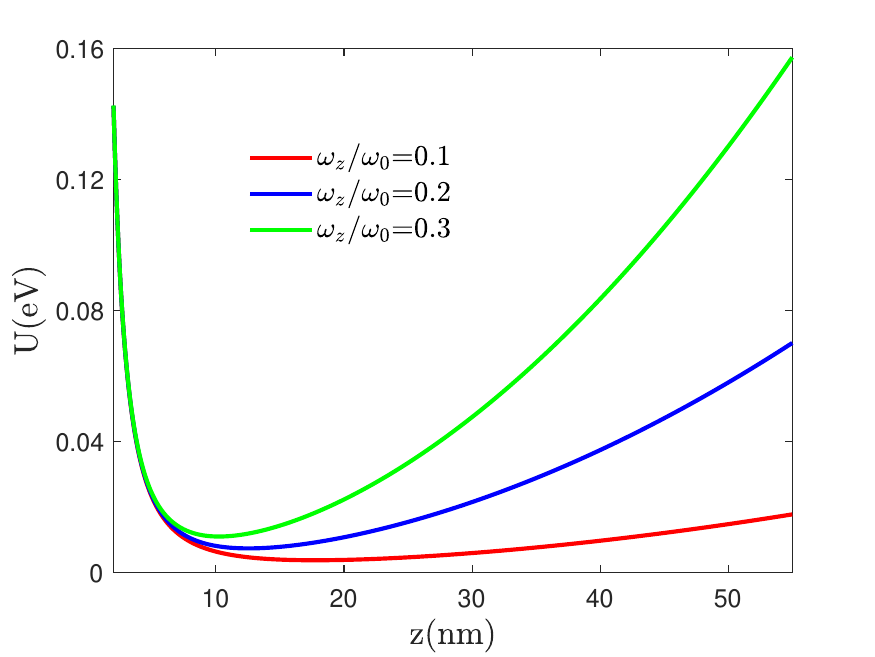}
    \caption{The potential form is displayed for three distinct values of ${{{\omega _z}} \mathord{\left/
 {\vphantom {{{\omega _z}} {{\omega _0}}}} \right.
 \kern-\nulldelimiterspace} {{\omega _0}}}$.}
    \label{Fig1}
\end{figure}

We consider a quantum well structure in which the electron moves freely in the XY plane and is confined along the z-axis by the confinement potential ISPPSISQW of the form
\begin{align}
    {U\left( z \right) = \left\{ {\begin{array}{*{20}{c}}
\infty &{z < 0}\\
{\dfrac{1}{2}{m_e}\omega _z^2{z^2} + \dfrac{{{\hbar ^2}{\beta _z}}}{{2{m_e}{z^2}}}}&{z > 0}
\end{array}} \right.}
\end{align}
in which, $m_e$ is an effective mass of the electron, $\hbar$ is reduced Planck's constant, $\beta_z$ and $\omega_z$ are the characteristic parameters of the potential well and the confinement frequency, respectively. The Quantum well model with the potential form is illustrated in Fig. \ref{Fig1} with different values of the confinement frequency $\omega _z$. 

The Schrödinger equation for electrons along the z-axis has the form
\begin{align}
    \left[ { - \dfrac{{{\hbar ^2}}}{{2{m_e}}}\dfrac{{{\partial ^2}}}{{\partial {z^2}}} + U\left( z \right)} \right]{\phi _{\rm{n}}}\left( z \right) = {\varepsilon _{\rm{n}}}{\phi _{\rm{n}}}\left( z \right)
\end{align}
Solving this equation we get wave functions and an energy spectrum of the electron in ISPPSISQW of the form \cite{lq1,lq2}
\begin{align}\label{wf}
    {{\phi _{\rm{n}}}\left( z \right) = {A_{\rm{n}}}{z^{2s}}\exp \left( { - \dfrac{{{z^2}}}{{2\alpha _z^2}}} \right){\cal L}_{\rm{n}}^\alpha \left( {\dfrac{{{z^2}}}{{\alpha _z^2}}} \right)}
\end{align}
\begin{align}\label{pnl}
    {\varepsilon _{\rm{n}}} = \left( {2{\rm{n}} + 1 + \dfrac{{\sqrt {1 + 4{\beta _z}} }}{2}} \right)\hbar {\omega _z}
\end{align}
Here, ${{\rm{s}} = \dfrac{1}{4}\left( {1 + \sqrt {1 + 4{\beta _z}} } \right)}$, ${{\alpha _z} = \sqrt {\dfrac{\hbar }{{{m_e}{\omega _z}}}} }$ and ${\cal L}_{\rm{n}}^{\alpha}(x)$ is the associated Laguerre polynomial. 

$A_{\rm{n}}$ is the wave function normalization coefficient \cite{lq1,lq2}
\begin{align}
{{A_{\rm{n}}} = \sqrt {\frac{{2{\rm{n}}!}}{{\alpha _z^{1 + 4{\rm{s}}}\Gamma \left( {2{\rm{s}} + {\rm{n}} + \frac{1}{2}} \right)}}} }
\end{align}
with $\Gamma(x)$ is the Gamma function. 

\subsection{Multiphoton nonlinear absorption coefficient in the case of the absence of an external magnetic field} 

When the SEMW is applied to the system, with the electric field vector ${\mathbf{E}} = \left( {0,{E_0}\sin \Omega t,0} \right)$ ($E_0$ and $\Omega$ are the amplitude and frequency, respectively), the Hamiltonian of the electron-phonon system in ISPPSISQW can be expressed in the second quantization representation as follows \cite{bau, tuan2023} 
\begin{align}
{\cal H} &= \displaystyle\sum\limits_{{\rm{n}},{{\bf{k}}_\bot}} {{\varepsilon _{\rm{n}}}\left( {{{\bf{k}}_ \bot } - \dfrac{e}{{\hbar c}}{\bf{A}}\left( t \right)} \right)a_{{\rm{n}},{{\bf{k}}_ \bot }}^\dag {a_{{\rm{n}},{{\bf{k}}_ \bot }}}}  + \displaystyle\sum\limits_{\bf{q}} {\hbar {\omega _{\bf{q}}} {b_{\bf{q}}^\dag {b_{\bf{q}}}} } \\
 &+ \displaystyle\sum\limits_{{\rm{n}},{\rm{n}}'} {\sum\limits_{{{\bf{k}}_ \bot },{\bf{q}}} {{\rm{C}}\left( {\bf{q}} \right){{\rm{J}}_{{\rm{n}}{\rm{,{\rm{n}}'}}}}\left( {\bf{q}} \right)a_{{\rm{n}}',{{\bf{k}}_ \bot } + {{\bf{q}}_ \bot }}^\dag {a_{{\rm{n}},{{\bf{k}}_ \bot }}}\left( {b_{ - {\bf{q}}}^\dag  + {b_{\bf{q}}}} \right)} } 
\end{align}
Here, $a_{{\rm{n}},{{\mathbf{k}}_{\bot}}}^ \dag $ and ${a_{{\rm{n}},{{\mathbf{k}}_{\bot}}}}$
are the creation and annihilation operators of electron (phonon), respectively, while $ {b_{\mathbf{q}}^ \dag \,{\text{and}}\,{b_{\mathbf{q}}}}$ are for the phonon; $\hbar {\omega _{\mathbf{q}}}$ is the phonon energy. The vector potential of laser radiation as a SEMW ${\mathbf{A}}\left( t \right) = \dfrac{c}{\Omega }{{\mathbf{E}}_0}\cos \left( {\Omega t} \right) $. ${\rm{n}}$ and ${\rm{n}}^{\prime}$ are the band indices of states $\left| {\bf{k}_ \bot} \right\rangle $ and $\left| {{\bf{k}_ \bot} + {\bf{q}}_ \bot} \right\rangle $, respectively. ${\rm{C}}\left( {\bf{q}} \right)$ is the electron-phonon interaction constant. We consider the case of electron interaction with dispersionless optical phonon $\hbar {\omega _{\bf{q}}} \approx \hbar {\omega _0}$ \cite{bau,bau4} 
\begin{align}
    {\left| {{\rm{C}}\left( {\bf{q}} \right)} \right|^2} = \dfrac{{2\pi {e^2}\hbar {\omega _0}}}{\epsilon_0{{{\bf{q}}^2}}}\left( {\dfrac{1}{{{\chi _\infty }}} - \dfrac{1}{{{\chi _0}}}} \right)
\end{align}
where ${\chi _\infty }$ and ${\chi _0}$ are the static and the high-frequency dielectric constants, respectively, and ${\bf{q}} = {{\bf{q}}_ \bot } + {{\bf{q}}_z}$ is the wave vector of the phonon. $\epsilon_0$ is the permittivity of the vacuum. 

The electron form factor is written as
\begin{align}\label{fo}
    {{\rm{J}}_{{\rm{n}},{\rm{n}}^\prime }}\left( {\bf{q}} \right) = {{\rm{J}}_{{\rm{n}},{\rm{n}}^\prime }}\left( {{q_z}} \right){\delta _{{{\bf{k}}_ \bot },{{\bf{k}}_ \bot } + {{\bf{q}}_ \bot }}} = \langle {\phi _{{\rm{n}}^\prime }}\left( z \right)|{{\rm{e}}^{i{q_z}z}}\left| {{\phi _{\rm{n}}}\left( z \right)} \right\rangle {\delta _{{{\bf{k}}_ \bot },{{\bf{k}}_ \bot } + {{\bf{q}}_ \bot }}}
\end{align}
Here, $\delta_{i,j}$ is the Kronecker Delta function. The integral in \eqref{fo} is calculated using a computer program. 

The quantum kinetic equation for electron distribution function \cite{bau, eps, eps1}  
\begin{align}\label{qke}
\dfrac{{\partial {f_{{\rm{n}},{{\mathbf{k}}_\bot}}}\left( t \right)}}{{\partial t}} =  - \dfrac{i}{\hbar }{\left\langle {\left[ {a_{{\rm{n}},{{\mathbf{k}}_\bot}}^ \dag {a_{{\rm{n}},{{\mathbf{k}}_\bot}}},{\mathcal{H}}} \right]} \right\rangle _t}
\end{align}
where ${f_{{\rm{n}},{{\mathbf{k}}_\bot}}}\left( t \right) = {\left\langle {a_{{\rm{n}},{{\mathbf{k}}_y}}^\dag {a_{{\rm{n}},{{\mathbf{k}}_\bot}}}} \right\rangle _t}$ is the electron distribution function, and  ${\left\langle ...  \right\rangle _t}$ denotes the statistical average value at the moment t. The solution of Eq. \eqref{qke} by first-order iterative approximation has the form \cite{bau, bau2}
\begin{align}
\begin{array}{l}
{f_{{\rm{n}},{{\bf{k}}_ \bot }}}\left( t \right) =  - \dfrac{1}{\hbar }\displaystyle\sum\limits_{{\bf{q}},{\rm{n}}^\prime } {{{\left| {{\rm{C}}\left( {\bf{q}} \right)} \right|}^2}{{\left| {{{\rm{J}}_{{\rm{n}},{\rm{n}}^\prime }}\left( {\bf{q}} \right)} \right|}^2}\displaystyle\sum\limits_{\ell ,\kappa } {{{\mathcal{J}}_\ell }\left( {\dfrac{{e{{\bf{E}}_0}{{\bf{q}}_ \bot }}}{{{m_e}{\Omega ^2}}}} \right){{\mathcal{J}}_{\ell  + \kappa }}\left( {\dfrac{{e{{\bf{E}}_0}{{\bf{q}}_ \bot }}}{{{m_e}{\Omega ^2}}}} \right)\dfrac{{{\rm{exp}}\left( { - i\kappa \Omega t} \right)}}{{\kappa \Omega }}} } \\
\left\{ { - \dfrac{{\overline {{f_{{\rm{n}},{{\bf{k}}_ \bot }}}} \overline {{{\rm{N}}_{\bf{q}}}}  - \overline {{f_{{\rm{n}}^\prime ,{{\bf{k}}_ \bot } + {{\bf{q}}_ \bot }}}} \left( {\overline {{{\rm{N}}_{\bf{q}}}}  + 1} \right)}}{{{\varepsilon _{{\rm{n}}^\prime ,{{\bf{k}}_ \bot } + {{\bf{q}}_ \bot }}} - {\varepsilon _{{\rm{n}},{{\bf{k}}_ \bot }}} - \hbar {\omega _0} - \ell \hbar \Omega  + i\hbar \delta }} - \dfrac{{\overline {{f_{{\rm{n}},{{\bf{k}}_ \bot }}}} \left( {\overline {{{\rm{N}}_{\bf{q}}}}  + 1} \right) - \overline {{f_{{\rm{n}}^\prime ,{{\bf{k}}_ \bot } + {{\bf{q}}_ \bot }}}} \overline {{{\rm{N}}_{\bf{q}}}} }}{{{\varepsilon _{{\rm{n}}^\prime ,{{\bf{k}}_ \bot } + {{\bf{q}}_ \bot }}} - {\varepsilon _{{\rm{n}},{{\bf{k}}_ \bot }}} + \hbar {\omega _0} - \ell \hbar \Omega  + i\hbar \delta }}} \right.\\
\left. { + \dfrac{{\overline {{f_{{\rm{n}}^\prime ,{{\bf{k}}_ \bot } - {{\bf{q}}_ \bot }}}} \overline {{{\rm{N}}_{\bf{q}}}}  - \overline {{f_{{\rm{n}},{{\bf{k}}_ \bot }}}} \left( {\overline {{{\rm{N}}_{\bf{q}}}}  + 1} \right)}}{{{\varepsilon _{{\rm{n}},{{\bf{k}}_ \bot }}} - {\varepsilon _{{\rm{n}}^\prime ,{{\bf{k}}_ \bot } - {{\bf{q}}_ \bot }}} - \hbar {\omega _0} - \ell \hbar \Omega  + i\hbar \delta }} + \dfrac{{\overline {{f_{{\rm{n}}^\prime ,{{\bf{k}}_ \bot } - {{\bf{q}}_ \bot }}}} \left( {\overline {{{\rm{N}}_{\bf{q}}}}  + 1} \right) - \overline {{f_{{\rm{n}},{{\bf{k}}_ \bot }}}} \overline {{{\rm{N}}_{\bf{q}}}} }}{{{\varepsilon _{{\rm{n}},{{\bf{k}}_ \bot }}} - {\varepsilon _{{\rm{n}}^\prime ,{{\bf{k}}_ \bot } - {{\bf{q}}_ \bot }}} + \hbar {\omega _0} - \ell \hbar \Omega  + i\hbar \delta }}} \right\}
\end{array}
\end{align}
where ${\mathcal{J}}_{\ell}(x)$ is the $\ell$th-order Bessel function of the argument $x$. ${\overline {{{\rm{N}}_{\bf{q}}}} }$ is the equilibrium distribution function for phonons, which is given by the Bose-Einstein distribution function. ${\overline {{f_{{\rm{n}},{{\bf{k}}_ \bot }}}} }$ is the equilibrium distribution function for electrons and be assumed to obey the Maxwell-Boltzmann distribution function \cite{bau}. $k_B$ is the Boltzmann constant, and T is the absolute temperature of the system. 

The total current density function has the form \cite{bau, eps, eps1} 
 \begin{align}\label{den}
\begin{array}{l}
{{\bf{J}}_ \bot }\left( t \right) = \dfrac{{e\hbar }}{{{m_e}}}\displaystyle\sum\limits_{{\rm{n}},{{\bf{k}}_ \bot }} {\left( {{{\bf{k}}_ \bot } - \dfrac{e}{{\hbar c}}{\bf{A}}\left( t \right)} \right){f_{{\rm{n}},{{\bf{k}}_ \bot }}}\left( t \right)} \\
 =  - \dfrac{{{e^2}{{\bf{E}}_0}{\eta _0}\cos \left( {\Omega t} \right)}}{{{m_e}\Omega }} + \dfrac{{2\pi e{\hbar ^2}}}{{{m_e}\kappa \Omega }}\displaystyle\sum\limits_{{\rm{n}},{\rm{n}}^\prime } {\sum\limits_{{{\bf{k}}_ \bot },{\bf{q}}} {{{\left| {{\rm{C}}\left( {\bf{q}} \right)} \right|}^2}{{\left| {{{\rm{J}}_{{\rm{n}},{\rm{n}}^\prime }}\left( {\bf{q}} \right)} \right|}^2}\overline {{{\rm{N}}_{\bf{q}}}} \left( {\overline {{f_{{\rm{n}},{{\bf{k}}_ \bot }}}}  - \overline {{f_{{\rm{n}}^\prime ,{{\bf{k}}_ \bot } + {{\bf{q}}_ \bot }}}} } \right)} } \\
 \times \displaystyle\sum\limits_\ell  {{{\bf{q}}_ \bot }{{\mathcal{J}}_\ell }\left( {\dfrac{{e{{\bf{E}}_0}{{\bf{q}}_ \bot }}}{{{m_e}{\Omega ^2}}}} \right)} \left[ {{{\mathcal{J}}_{\ell  + \kappa }}\left( {\dfrac{{e{{\bf{E}}_0}{{\bf{q}}_ \bot }}}{{{m_e}{\Omega ^2}}}} \right) + {{\mathcal{J}}_{\ell  - \kappa }}\left( {\dfrac{{e{{\bf{E}}_0}{{\bf{q}}_ \bot }}}{{{m_e}{\Omega ^2}}}} \right)} \right]\delta \left( {{\varepsilon _{{\rm{n}}^\prime ,{{\bf{k}}_ \bot } + {{\bf{q}}_ \bot }}} - {\varepsilon _{{\rm{n}},{{\bf{k}}_ \bot }}} + \hbar {\omega _0} - \ell \hbar \Omega } \right)
\end{array}
 \end{align}
 Here, ${\eta _0} = \displaystyle\sum\limits_{{\rm{n}},{{\bf{k}}_ \bot }} {{f_{{\rm{n}},{{\bf{k}}_ \bot }}}\left( t \right)} $ is the electron density in ISPPSISQW. $\delta(x)$ is the Dirac Delta function

From the expression for the total current density vector in Eq. \eqref{den}, we obtain the  total absorption coefficient (TAC) of the electromagnetic wave in ISPPSISQW \cite{eps1, bau, bau1}
\begin{align}\label{a}
\begin{array}{l}
\alpha  = \dfrac{{8\pi }}{{c\sqrt {{\chi _\infty }} E_0^2}}{\left\langle {{{\bf{J}}_ \bot } \cdot {\bf{E}}} \right\rangle _t} = \dfrac{{2{m_e}{\eta _0}{e^2}\Omega {k_B}T}}{{c\sqrt {{\chi _\infty }} E_0^2{\hbar ^2}{\varepsilon _0}\pi }}\left( {\dfrac{1}{{{\chi _\infty }}} - \dfrac{1}{{{\chi _0}}}} \right)\displaystyle\sum\limits_{{\rm{n}},{\rm{n}}^\prime } {\left| {{{\rm{G}}_{{\rm{n}},{\rm{n}}^\prime }}} \right|{\rm{exp}}\left( {\dfrac{{{\varepsilon _{\rm{F}}} - {\varepsilon _{\rm{n}}}}}{{{k_B}T}}} \right)} {\Im _\ell }\\
{{\rm{G}}_{{\rm{n}},{\rm{n}}^\prime }} = \displaystyle\int\limits_{ - \infty }^{ + \infty } {{{\left| {{{\rm{J}}_{{\rm{n}},{\rm{n}}^\prime }}\left( {{q_z}} \right)} \right|}^2}d{q_z}}  \\
{\Im _\ell } = \displaystyle\sum\limits_\ell  {\dfrac{\ell }{{{{\left( {\ell !} \right)}^2}}}{{\left( {\dfrac{{e{E_0}}}{{2{m_e}{\Omega ^2}}}} \right)}^{2\ell }}} {\rm{exp}}\left( { - \dfrac{{{{\rm{D}}_{{\rm{n}}^\prime ,{\rm{n}}}}}}{{2{k_B}T}}} \right){\left[ {\dfrac{{2{m_e}}}{{{\hbar ^2}}}{{\rm{D}}_{{\rm{n}}^\prime ,{\rm{n}}}}} \right]^\ell }{{\bf{K}}_\ell }\left( {\dfrac{{{{\rm{D}}_{{\rm{n}}^\prime ,{\rm{n}}}}}}{{2{k_B}T}}} \right)\\
{{\rm{D}}_{{\rm{n}}^\prime ,{\rm{n}}}} = {\varepsilon _{{\rm{n}}^\prime }} - {\varepsilon _{\rm{n}}} + \hbar {\omega _0} - \ell \hbar \Omega 
\end{array}
\end{align}
where $c$ is the speed of light in a vacuum and $\varepsilon_{\rm{F}}$ is Fermi energy. ${{\bf{K}}_\ell }\left( x \right)$ is modified Bessel function of the second kind. The overlap integral due to electron–phonon interaction in \eqref{a} is calculated using computer programs. ${\Im _\ell }$ is the characteristic term for the $\ell$-photon absorption process, with $\ell = 1$ and 2 corresponding to the linear absorption process and the nonlinear absorption process. 

\subsection{Multiphoton nonlinear absorption coefficient in the case of the presence of an external magnetic field}
In this section, we consider a semiconductor specimen subjected to an external magnetic field perpendicular to it. The external magnetic field is parallel to the confinement axis Oz $\mathbf{B}= \left(0,0, B\right) $ with vector potential selected according to Landau gauge of the form $\mathbf{A}^{\prime} = \left( {0, Bx,0} \right)$. The wave function and the corresponding energy are written as 
\begin{align}\label{wfb}
    \Psi \left( {\bf{r}} \right) = \left| {{\rm{N}},{\rm{n}},{{\bf{k}}_y}} \right\rangle  = \frac{1}{{\sqrt {{L_y}} }}\exp \left( {i{k_y}y} \right){\Phi _{\rm{N}}}\left( {x - {x_0}} \right){\phi _{\rm{n}}}\left( z \right)
\end{align}
\begin{align}\label{pnlb}
    {{\varepsilon _{{\rm{N}},{\rm{n}}}} = {\varepsilon _{\rm{N}}} + {\varepsilon _{\rm{n}}} = \left( {{\rm{N}} + \frac{1}{2}} \right)\hbar {\omega _{\rm{B}}} + {\varepsilon _{\rm{n}}}}
\end{align}
here, ${\phi _{\rm{n}}}\left( z \right)$ and ${{\varepsilon _{\rm{n}}}}$ have been shown in Eqs. \eqref{wf} and \eqref{pnl}. ${\rm{N}} = 0,1,2,...$ is Landau indices. ${{\omega _{\rm{B}}} = \dfrac{{eB}}{{{m_{e}c}}}}$ is the cyclotron frequency and  ${\Phi _{\rm{N}}}\left( {x - {x_0}} \right)$ is the normalized harmonic oscillator function  
\begin{align}
    {\Phi _{\rm{N}}}\left( {x - {x_0}} \right) = \left| {\rm{N}} \right\rangle  = \sqrt {\frac{1}{{{2^{\rm{N}}}{\rm{N}}!\sqrt \pi  {\ell _{\rm{B}}}}}} \exp \left[ { - \frac{{{{\left( {x - {x_0}} \right)}^2}}}{{\ell _{\rm{B}}^2}}} \right]{{\rm H}_{\rm{N}}}\left( {\frac{{x - {x_0}}}{{{\ell _{\rm{B}}}}}} \right)
\end{align}
with ${{{\rm H}_{\rm{N}}}\left( x \right) = {{\left( { - 1} \right)}^{\rm{N}}}\exp \left( {{x^2}} \right)\dfrac{{{{\rm{d}}^{\rm{N}}}}}{{{\rm{d}}{x^{\rm{N}}}}}\left[ {\exp \left( { - {x^2}} \right)} \right]}$ is the N-th order Hermite polynomial. ${x_0}$ being the coordinate of the center of the carrier orbit ${{x_0} =  - \ell _{\rm{B}}^2{k_y}}$, in which, ${{\ell _{\rm{B}}} = \sqrt {\dfrac{{c\hbar }}{{eB}}} }$ being the magnetic length. 

Interaction Hamiltonian of the electron-optical phonon system in the presence of a magnetic field of the form
\begin{align}\label{hint}
{\cal H} &= \sum\limits_{{\rm{N}},{\rm{n}},{{\bf{k}}_y}} {{\varepsilon _{{\rm{N}},{\rm{n}},{{\bf{k}}_y}}}\left[ {{{\bf{k}}_y} - \frac{e}{{c\hbar }}{\bf{A}}\left( t \right)} \right]a_{{\rm{N}}^\prime ,{\rm{n}}^\prime ,{{\bf{k}}_y} + {{\bf{q}}_y}}^\dag {a_{{\rm{N}},{\rm{n}},{{\bf{k}}_y}}}}  + \sum\limits_{\bf{q}} {\hbar {\omega _{\bf{q}}}b_{\bf{q}}^\dag {b_{\bf{q}}}} \\
 &+ \sum\limits_{{\rm{N}}^\prime ,{\rm{N}}} {\sum\limits_{{\rm{n}}^\prime ,{\rm{n}}} {\sum\limits_{{{\bf{k}}_y},{\bf{q}}} {{\rm{C}}\left( {\bf{q}} \right){{\rm{J}}_{{\rm{n}}^\prime ,{\rm{n}}}}\left( {\bf{q}} \right){{\rm I}_{{\rm{N}}^\prime ,{\rm{N}}}}\left( {{{\bf{q}}_ \bot }} \right)a_{{\rm{N}}^\prime ,{\rm{n}}^\prime ,{{\bf{k}}_y} + {{\bf{q}}_y}}^\dag {a_{{\rm{N}},{\rm{n}},{{\bf{k}}_y}}}\left( {b_{ - {\bf{q}}}^\dag  + {b_{\bf{q}}}} \right)} } } 
\end{align}

${{\rm I}_{{\rm{N}}^\prime,{\rm{N}}}}\left( {\bf{q}} \right)$ is the electron form factor under the influence of the magnetic field \cite{ryu1,ryu2}
\begin{align}\label{tsd}
    {{\rm I}_{{\rm{N}}^\prime ,{\rm{N}}}}\left( {{{\bf{q}}_ \bot }} \right) = \left\langle {{\Phi _{{\rm{N}}^\prime }}\left( {x - {x_0}} \right)} \right|\exp \left( {i{q_x}x} \right)\left| {{\Phi _{\rm{N}}}\left( {x - {x_0}} \right)} \right\rangle {\delta _{{{\bf{k}}_y},{{\bf{k}}_y} + {{\bf{q}}_y}}}
\end{align}

To determine the integral in Eq. \eqref{tsd}, we set $u = \dfrac{{\ell _{\rm{B}}^2{\bf{q}}_ \bot ^2}}{2}$ and after some calculations similar to the previous works \cite{ryu1,ryu2}, we obtain the electron form factor as follows \cite{enck1969phonon}
\begin{align}
    {\left| {{{\rm I}_{{\rm{N}}^\prime ,{\rm{N}}}}\left( u \right)} \right|^2} = \frac{{{\rm{N}}!}}{{{\rm{N}}^\prime !}}\exp \left( { - u} \right){u^{{\rm{N}}^\prime  - {\rm{N}}}}{\cal L}_{\rm{N}}^{{\rm{N}}^\prime  - {\rm{N}}}\left( u \right)
\end{align}

Using a technique similar to that used to solve Eq. \eqref{qke}, we solve the quantum kinetic equation, which is established for electrons in ISPPSISQW for the case of an external magnetic field, and obtain the TAC 
\begin{align}\label{alb}
\alpha  = \frac{{8{\pi ^3}{e^4}{k_B}T{\eta _0}}}{{c\sqrt {{\chi _\infty }} {\varepsilon _0}m_e^2\ell _B^2{\Omega ^3}}}\left( {\frac{1}{{{\chi _\infty }}} - \frac{1}{{{\chi _0}}}} \right)\displaystyle\sum\limits_{{{\rm{N}}^\prime },{\rm{N}}} {\sum\limits_{{{\rm{n}}^\prime },{\rm{n}}} {{\Lambda _{{\rm{N}},{\rm{n}}}}\left| {{{\rm{G}}_{{{\rm{n}}^\prime },{\rm{n}}}}} \right|} } \sum\limits_\ell  {{{\cal A}_{\ell }}\delta \left( {{\varepsilon _{{{\rm{N}}^\prime },{{\rm{n}}^\prime }}} - {\varepsilon _{{\rm{N}},{\rm{n}}}} + \hbar {\omega _0} - \ell \hbar \Omega } \right)} 
\end{align}
with ${\Lambda _{{\rm{N}},{\rm{n}}}} = {{\exp \left( {\displaystyle\frac{{{\varepsilon _{\rm{F}}} - {\varepsilon _{{\rm{N}},{\rm{n}}}}}}{{{k_B}T}}} \right)} \mathord{\left/
 {\vphantom {{\exp \left( {\displaystyle\frac{{{\varepsilon _{\rm{F}}} - {\varepsilon _{{\rm{N}},{\rm{n}}}}}}{{{k_B}T}}} \right)} {\sum\limits_{{\rm{N}},{\rm{n}}} {\exp \left( {\displaystyle\frac{{{\varepsilon _{\rm{F}}} - {\varepsilon _{{\rm{N}},{\rm{n}}}}}}{{{k_B}T}}} \right)} }}} \right.
 \kern-\nulldelimiterspace} {\displaystyle\sum\limits_{{\rm{N}},{\rm{n}}} {\exp \left( {\displaystyle\frac{{{\varepsilon _{\rm{F}}} - {\varepsilon _{{\rm{N}},{\rm{n}}}}}}{{{k_B}T}}} \right)} }}$ and ${{\cal A}_{\ell }}$ is the dimensionless parameter characterizing $\ell$-photon absorption process. For mono-photon absorption (linear absorption process), ${{\cal A}_1} = 1$. For two-photon absorption (nonlinear absorption process) ${{\cal A}_2} = {{{\left( {\dfrac{{e{E_0}}}{{2{m_e}{\Omega ^2}}}} \right)}^2}\left( {{\rm{N}} + {{\rm{N}}^\prime } + 1} \right)}$. The Dirac delta functions appearing in Eq. \eqref{alb} will diverge as the argument approaches 0. To avoid this, we use the Lorentzian transformation proposed by C. M. Van Vliet as follows \cite{van1,van3} 
\begin{align}
\delta \left( {{\varepsilon _{{\rm{N}}^\prime ,{\rm{n}}^\prime }} - {\varepsilon _{{\rm{N}},{\rm{n}}}} + \hbar {\omega _0} - \ell \hbar \Omega } \right) \Rightarrow \frac{1}{\pi }\frac{{\hbar {\gamma _{{\rm{N}},{\rm{n}}{{\rm{,N}},{\rm{N}}^\prime }}}}}{{{{\left( {{\varepsilon _{{\rm{N}}^\prime ,{\rm{n}}^\prime }} - {\varepsilon _{{\rm{N}},{\rm{n}}}} + \hbar {\omega _0} - \ell \hbar \Omega } \right)}^2} + {\hbar ^2}\gamma _{{\rm{N}},{\rm{n}}{{\rm{,N}},{\rm{N}}^\prime }}^2}}
\end{align}
Here, $\gamma _{{\rm{n}},{\rm{n}}^\prime {\rm{, N}},{\rm{N}}^\prime }$ is the inverse relaxation time of the electron, which is given by the following expression 
\begin{align}
    {\left( {\gamma _{{\rm{n}},{{\rm{n}}^\prime },{\rm{N}},{{\rm{N}}^\prime }}^ \pm } \right)^2} = \frac{1}{{{\hbar ^2}}}\sum\limits_{\bf{q}} { {{\left| {{\rm{C}}\left( {\bf{q}} \right)} \right|}^2}{{\left| {{{\rm{J}}_{{\rm{n}},{{\rm{n}}^\prime }}}\left( {\bf{q}} \right)} \right|}^2}{{\left| {{{\rm{I}}_{{\rm{N}},{{\rm{N}}^\prime }}}\left( {\bf{q}} \right)} \right|}^2}\left( {\overline {{{\rm{N}}_{\bf{q}}}}  + \frac{1}{2} \pm \frac{1}{2}} \right)}  
\end{align}

The analytic expressions of the general TAC in the presence of a magnetic field are extremely complicated. In the next section, using numerical programs, we will plot and show the dependence of TAC on important parameters of the system and the external field. In addition, we also provide a more detailed discussion of the physical significance of the analytic results obtained here. 

\section{Numerical results and discussions}
In this section, we numerically calculated the TAC for the specific case of a GaAs/GaAsAl ISPPSISQW. The parameters used in the calculations are as follows \cite{bau,bau4, phong} ${\chi _\infty } = 10.9;{\chi _0} = 13.1;{m_e} = 0.067{m_0};{\eta _0} = {10^{ - 23}}{{\rm{m}}^{{\rm{ - 3}}}};\hbar {\omega _0} = 36.25{\rm{meV}}$ and $\varepsilon_{\rm{F}} = 50 \rm{meV}$, $m_0$ being the mass of free
electron. 

\subsection{In the Absence of an External Magnetic Field}
\begin{figure}[!htb]
\centering
\subfigure[][\label{1a}]
  {\includegraphics[width=0.47\linewidth]{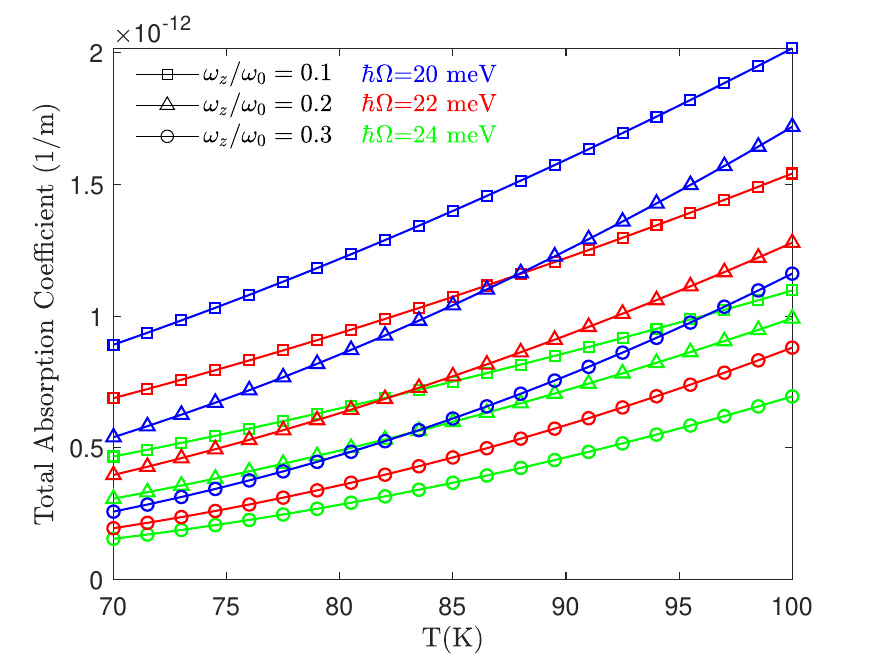}}
\subfigure[][\label{1b}]
  {\includegraphics[width=0.47\linewidth]{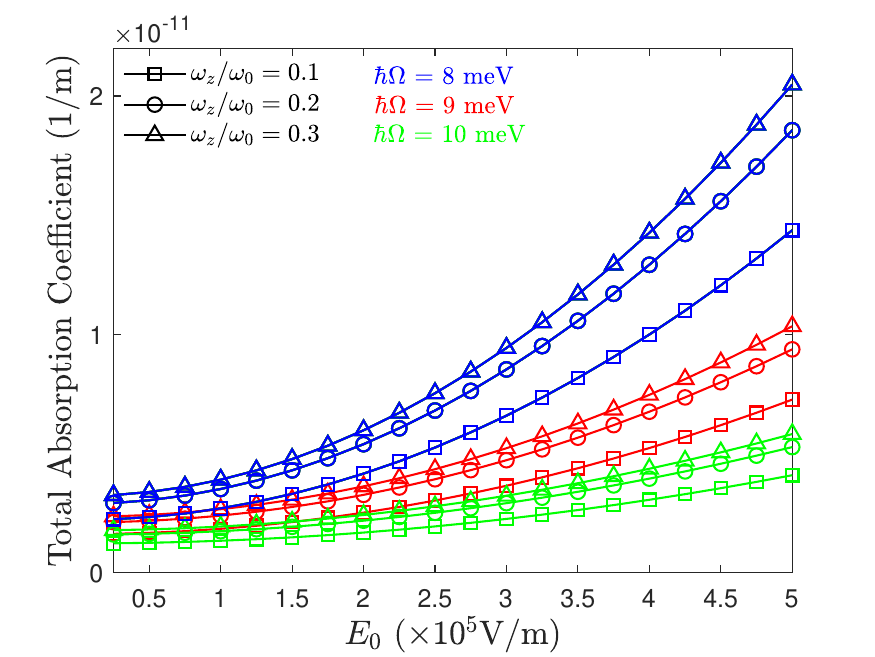}}
\caption{(Color online) TAC as a function of the temperature (left) and the intensity of SEMW (right) at three different values of the photon energy and the ratio ${{{\omega _z}} \mathord{\left/
 {\vphantom {{{\omega _z}} {{\omega _0}}}} \right.
 \kern-\nulldelimiterspace} {{\omega _0}}}$.}
\label{tacet}
\end{figure}

In Fig. \ref{1a}, the graph shows the increase of TAC as the temperature of the system increases with different photon energy values of the SEMW. This is consistent with the results obtained in bulk semiconductors, where increasing temperature leads to an increase in the TAC \cite{eps}. However, this result is contrary to previous studies on other two-dimensional electron gas systems \cite{bau, phong}.     

From Fig. \ref{1b}, we can see that the dependence of the TAC on the intensity $E_0$ is nonlinear. The TAC grows as the amplitude $E_0$ of the SEMW increases; however, as the photon energy increases, their TAC diminishes. These results are also true in the case of absorption of previously investigated one-dimensional electronic systems such as quantum wires \cite{bau1,phong1}, which shows a significant influence of strong electromagnetic waves on low-dimensional semiconductor systems. 

\begin{figure}[!htb]
\centering
\subfigure[][Three different values of the temperature.\label{2a}]
  {\includegraphics[width=0.47\linewidth]{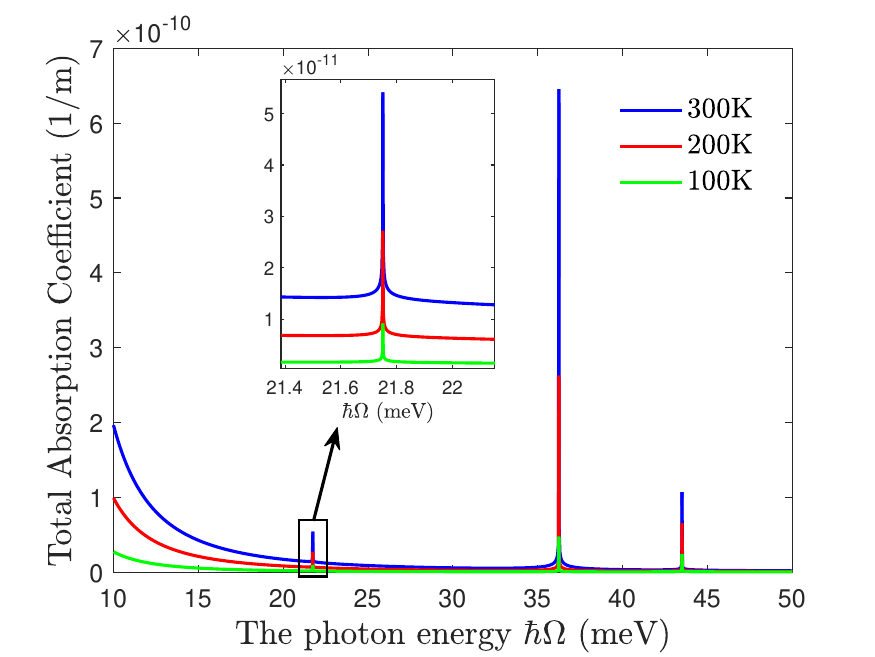}}
\subfigure[][Three different values of the ratio ${{{\omega _z}} \mathord{\left/
 {\vphantom {{{\omega _z}} {{\omega _0}}}} \right.
 \kern-\nulldelimiterspace} {{\omega _0}}}$. \label{2b}]
  {\includegraphics[width=0.47\linewidth]{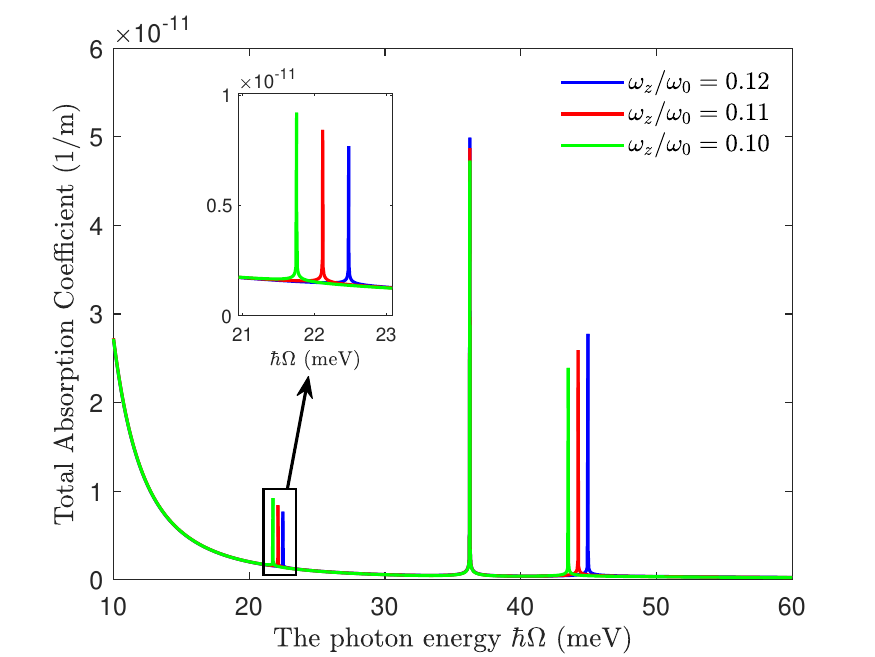}}
\caption{(Color online) TAC as a function of the photon energy of electromagnetic wave.}
\label{sosanhdischo}
\end{figure}

In Fig. \ref{2a}, we show the dependence of the TAC on the photon energy at different values of the temperature. We consider the electron's transition from ground state $\left| {{\rm{n}}} \right\rangle  = \left| {0} \right\rangle $ to first exited state $\left| {{\rm{n}}^\prime } \right\rangle  = \left| {1} \right\rangle $ and intrasubband transition ${\rm{n}} \to {\rm{n}}^\prime  = {\rm{n}}$. We can see the appearance of optical absorption spectral lines whose position is independent of temperature. Indeed, from the general expressions, the electron-phonon resonance condition (EPRC) can be obtained by setting the arguments in delta functions to zeros. The maximum peaks appear satisfying the EPRC ${\varepsilon _{{\rm{n}}^\prime }} - {\varepsilon _{\rm{n}}} + \hbar {\omega _0} - \ell \hbar \Omega  = 0$ or in other words, from Eq. \eqref{pnl}, we have $2\hbar {\omega _z}\left( {{{\rm{n}}^\prime } - {\rm{n}}} \right) + \hbar {\omega _0} = \ell \hbar \Omega $. Specifically, we only consider the transitions from $\rm{n} = 0$ to $\rm{n}^{\prime} = 1$, we can see that the first maximum peak appears at position $\hbar\Omega \approx 21.75$ meV satisfying the EPRC due to the two-photon absorption process. The mid-maximal peak occurs when the electromagnetic wave photon energy is equal to the optical phonon energy emitted by the intraband transition $\hbar \Omega  = \hbar {\omega _0} = 36.25{\rm{meV}}$. The last peak appears at position $\hbar\Omega \approx 43.5$ meV satisfying the EPRC due to the one-photon absorption process. In addition, from the EPRC, we also show that the position of these peaks is independent of temperature \cite{bau,bau1}. This suggests that temperature is not an important factor in the optical detection of the electric subbands. 

In Fig. \ref{2b}, we show the dependence of the TAC on the photon energy at different values of the ratio ${{{\omega _z}} \mathord{\left/
 {\vphantom {{{\omega _z}} {{\omega _0}}}} \right.
 \kern-\nulldelimiterspace} {{\omega _0}}}$. The appearance of the resonance peaks in this figure still basically obeys the EPRC condition. As the confinement frequency ${{\omega _z}}$ increases, the resonance peak position is shifted to the right. Notably, the position of the maximum peak due to the intraband transition $\hbar \Omega  = \hbar {\omega _0} = 36.25{\rm{meV}}$ does not depend on the value of the confinement frequency ${{\omega _z}}$. This can be explained based on EPRC as follows. Substituting Eq. \eqref{pnl} into EPRC, we have $\ell \hbar \Omega  = 2\hbar {\omega _z}\left( {{\rm{n}}^\prime  - {\rm{n}}} \right) + \hbar {\omega _0}$. According to this relation, when we consider the transitions from $\rm{n} = 0$ to ${{\rm{n}}^\prime } = 1$, if ${{\omega _z}}$ increases, ${{\hbar\Omega}}$ also increases, leading to a right shift of the resonance peaks in Fig. \ref{2b}. In addition, if we consider the intraband transitions, when $\rm{n} = {{\rm{n}}^\prime }$, the value of ${{\omega _z}}$ does not affect the value of ${{\hbar\Omega}}$, representing the position of the resonance peaks of the intraband transitions. Besides, we also find that as ${{\omega _z}}$ increases, the confinement potential profile decreases (see Fig. \ref{Fig1}), leading to an increase in the scattering probability between electrons and optical phonons, causing the TAC to increase. This finding is consistent with previous research in quantum well systems with various confinement potential profiles \cite{van, tung,lvtung,phuc2015,tung2}. However, the effect of ${{\omega _z}}$ on the Full Width at Half Maximum (FWHM) is not significant. This is shown in Figure 4a, illustrating the dependence of the FWHM of linear absorption coefficient (FWHM-LAC) and nonlinear absorption coefficient (FWHM-NAC) on the confinement frequency ${{\omega _z}}$. We also observe that the value of FWHM-LAC is larger than that of FWHM-NAC. Besides, the graph in Fig. \ref{4b} shows that both FWHM-LAC and FWHM-NAC increase with increasing temperature. Because FWHM is closely related to the rate of total momentum relaxation time, it depends on the specific physical properties of the scattering mechanism, of which temperature is one. It follows that as the temperature increases, the scattering probability increases and thus the FWHM also increases.

\subsection{In the Presence of an External Magnetic Field}

The graph in Fig. \ref{5a} shows that the dependence of LAC and NAC on the temperature of the system is a monotonically increasing function. The value of LAC (absorbs 1 photon) is greater than the value of NAC (absorbs 2 photons). As the confinement frequency $\omega_z$ increases, the potential profile decreases, causing the scattering probability to increase, causing the absorption coefficient to increase. Fig. \ref{5b} shows the dependence of LAC and NAC on the intensity of the SEMW. The LAC value does not depend on the intensity of the SEMW (because this is linear absorption). 

We only consider the electron's transition from ground state $\left| {{\rm{N}},{\rm{n}}} \right\rangle  = \left| {0,0} \right\rangle $ to first exited state $\left| {{\rm{N}}^\prime ,{\rm{n}}^\prime } \right\rangle  = \left| {1,1} \right\rangle $. The graphs examine the dependence of TAC on the photon energy of the SEMW, the temperature T and external magnetic field B are shown in Figs. \ref{fig6}, \ref{fig7}, and \ref{fig8}. The effects of temperature and photon energy of SEMW on TAC are shown in Figs. \ref{6a} and \ref{7a}. Temperature does not affect the position of the resonance peaks of the electron's intersubband transitions, but the greater the temperature, the higher the resonance peak. From Figs. \ref{6a} and \ref{7a}, under the influence of an external magnetic field B = 16T, one and two-photon absorption peaks appear at positions $\hbar \Omega  \approx 39.2$meV and $\hbar \Omega  \approx 78.4$meV. The position of the peak obeys the magneto-phonon resonance condition (MPRC) $\ell \hbar \Omega  = {\varepsilon _{{{\rm{N}}^\prime },{{\rm{n}}^\prime }}} - {\varepsilon _{{\rm{N}},{\rm{n}}}} + \hbar {\omega _0} = \hbar {\omega _B}\left( {{\rm{N}}^\prime  - {\rm{N}}} \right) + 2\hbar {\omega _z}\left( {{\rm{n}}^\prime  - {\rm{n}}} \right) + \hbar {\omega _0}$. When ${\rm{n}} =0$, ${\rm{n}}^{\prime} =1$ and ${\rm{N}} =0$, ${\rm{N}}^{\prime} =1$, MPRC becomes 
$\ell \hbar \Omega  = \hbar {\omega _B} + \hbar {\omega _z} + \hbar {\omega _0}$. Using this result, we can easily determine the location of resonance peaks for each different value of the magnetic field. We also find that temperature T does not appear in the MPRC, which explains why temperature does not affect the position of the resonance peaks as shown in Fig. \ref{6a}. For a more general view, in Fig. \ref{7b}, we show the density plot of the TAC versus temperature and photon energy of the SEMW. We can see that as the temperature increases the width of magneto-phonon resonance peaks at the same value of TAC increase. We will examine in detail the dependence of the width of the resonance peaks on temperature in Fig. \ref{9b}. 

Additionally, according to MPRC, as the external magnetic field increases, the effective magnetic energy $\hbar\omega_{B}$ increases (recall, $\hbar {\omega _B} = {{eB} \mathord{\left/
 {\vphantom {{eB} {\left( {{m_e}c} \right)}}} \right.
 \kern-\nulldelimiterspace} {\left( {{m_e}c} \right)}}$), which leads to a rightward shift of the resonance peaks as shown in Fig. \ref{6b}. However, from Fig. \ref{8a}, we see that the magnitude value of TAC increases with the magnetic field in the range ${\rm{B}} <  8.6 {\rm{T}}$ (see also Fig. \ref{6b}). TAC reaches its maximum value when the external magnetic field B = 8.6T and decreases as the magnetic field continues to increase. The graph showing the dependence of the density of TAC on the magnetic field and photon energy of the SEMW in Fig. \ref{8b} illustrates this statement more clearly. Besides, the numerical calculation results also show that the linear absorption peak (absorbing 1 photon) has a value 2.2 times larger than the nonlinear absorption peak (absorbing 2 photons). In other words, the magnitude value of NAC (two-photon absorption process) is 45.4 $\%$ of the magnitude value of LAC (one-photon absorption process). Therefore, the process of absorbing more than 1 photon (multi-photon absorption) is significant and cannot be ignored in the theoretical study on the absorption effect of SEMW in low-dimensional semiconductor systems in general. This result is consistent with previous research \cite{tung} using perturbation theory methods. 

\begin{figure}[!htb]
\centering
\subfigure[][\label{4a}]
  {\includegraphics[width=0.47\linewidth]{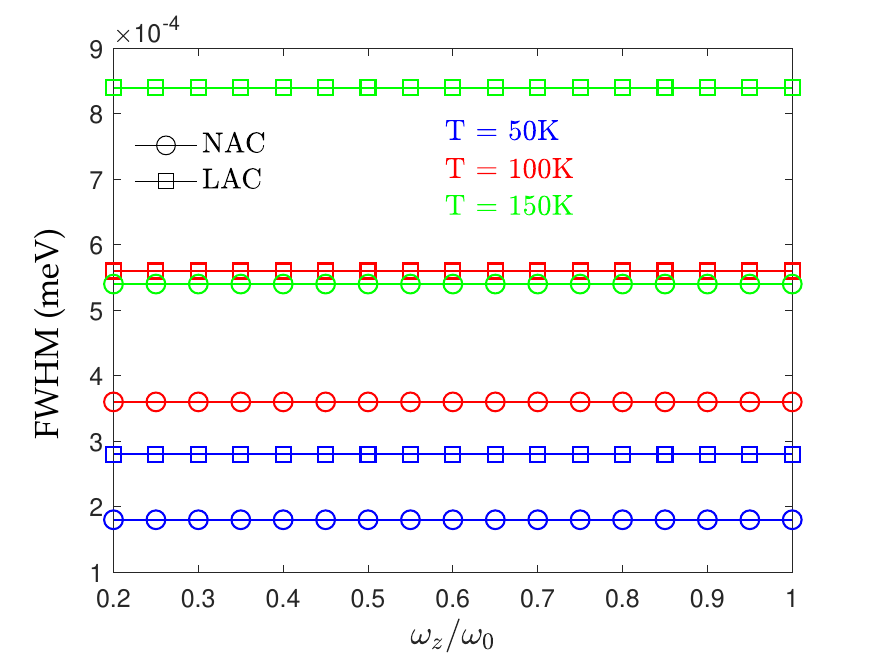}}
\subfigure[][ \label{4b}]
  {\includegraphics[width=0.47\linewidth]{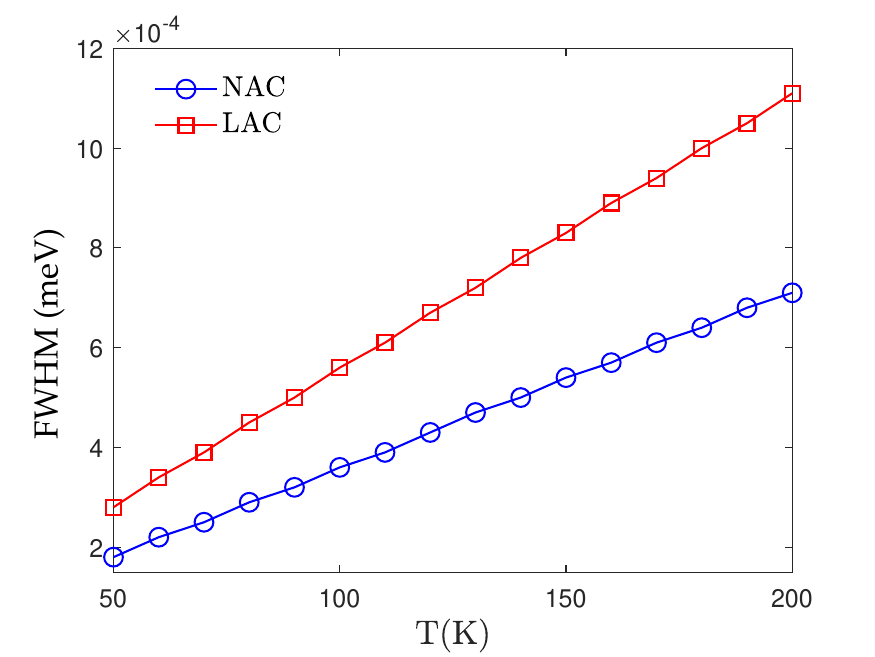}}
\caption{(Color online) FWHM of Linear Absorption Coefficient (LAC) (empty squares) and Nonlinear Absorption Coefficient (NAC) (empty circles) as a function of the ratio ${{{\omega _z}} \mathord{\left/
 {\vphantom {{{\omega _z}} {{\omega _0}}}} \right.
 \kern-\nulldelimiterspace} {{\omega _0}}}$ (left) and the temperature (right).}
\label{hwhmkob}
\end{figure}

\begin{figure}[!htb]
\centering
\subfigure[][\label{5a}]
  {\includegraphics[width=0.47\linewidth]{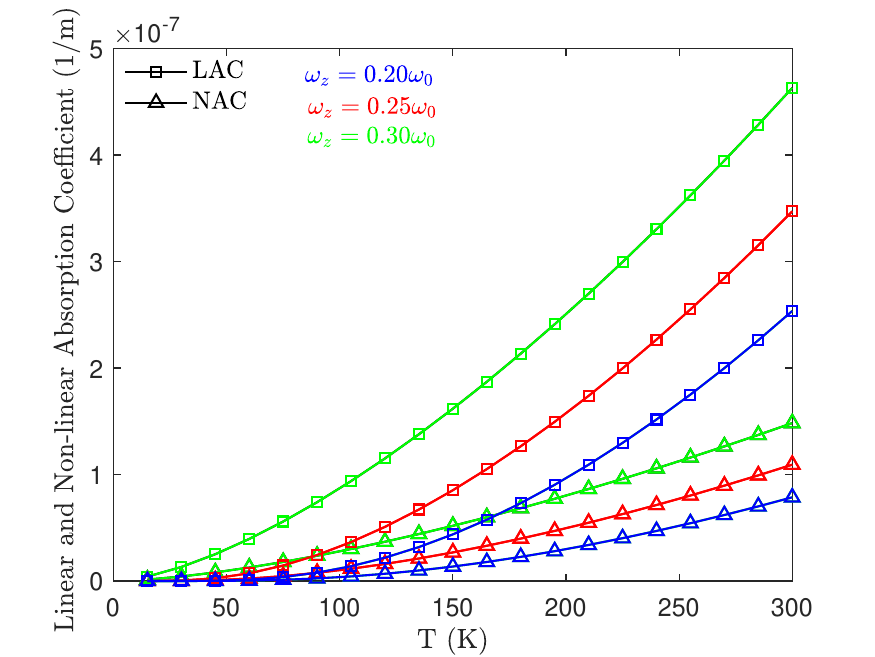}}
\subfigure[][\label{5b}]
  {\includegraphics[width=0.47\linewidth]{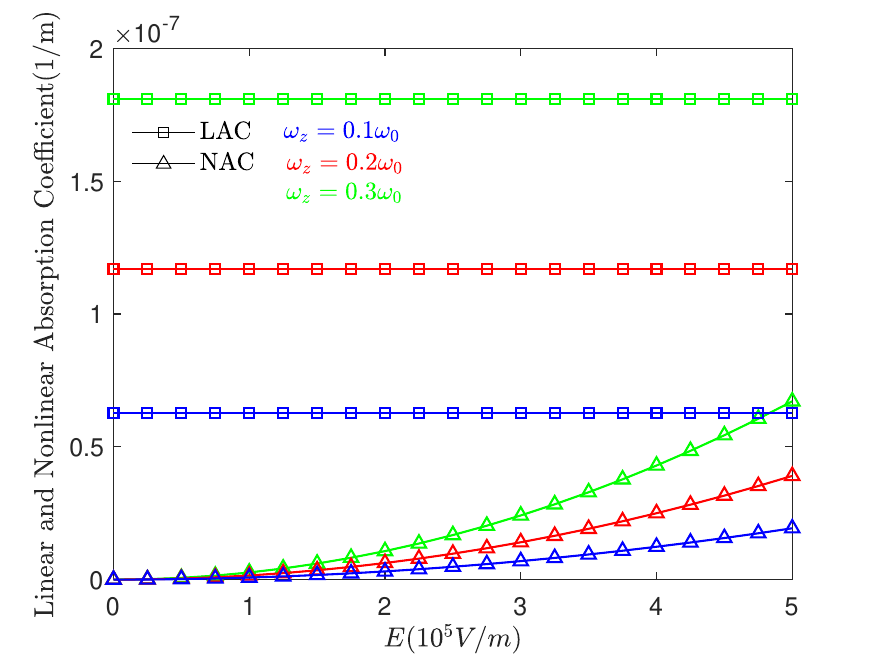}}
\caption{(Color online) Linear Absorption Coefficient (LAC) and Nonlinear Absorption Coefficient (NAC) as a function of the temperature (left) and the intensity of SEMW (right) at three different values of the ratio ${{{\omega _z}} \mathord{\left/
 {\vphantom {{{\omega _z}} {{\omega _0}}}} \right.
 \kern-\nulldelimiterspace} {{\omega _0}}}$ in the presence of magnetic field with a magnitude of B = 16T.}
\label{fig5}
\end{figure}

\begin{figure}[!htb]
\centering
\subfigure[][$B=16$T, $\omega_z=0.1\omega_0$\label{6a}]
  {\includegraphics[width=0.47\linewidth]{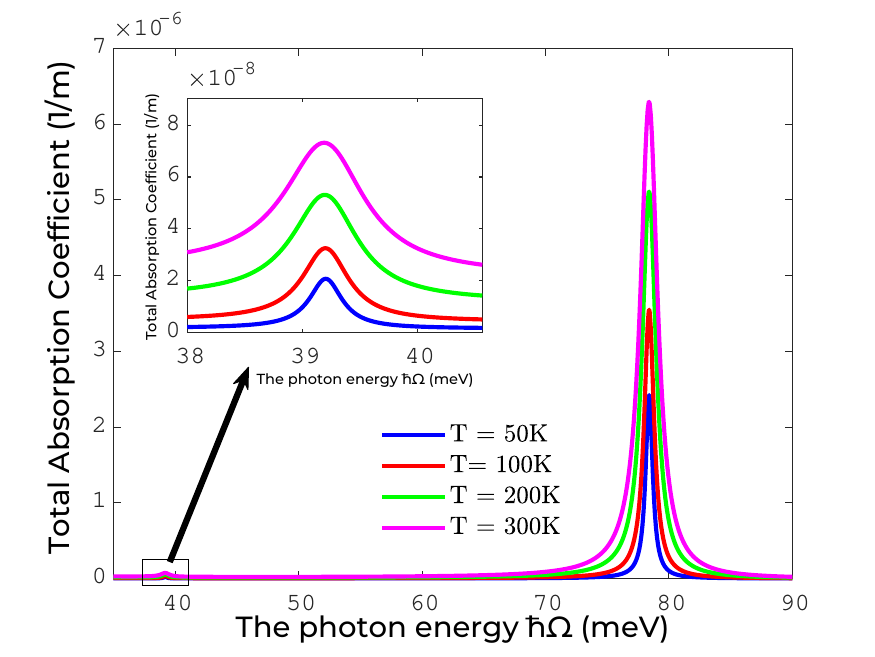}}
\subfigure[][$\omega_z=0.1\omega_0$\label{6b}]
  {\includegraphics[width=0.47\linewidth]{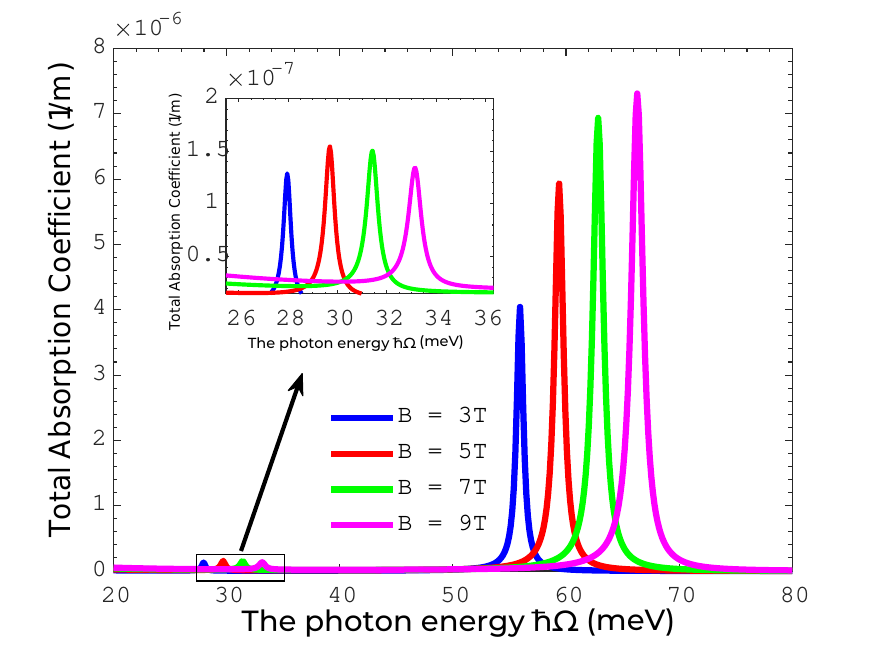}}
\caption{(Color online) TAC as a function of the photon energy with four different values of the temperature (left) and the magnetic field (right).}
\label{fig6}
\end{figure}

\begin{figure}[!htb]
\centering
\subfigure[][\label{7a}]
  {\includegraphics[width=0.47\linewidth]{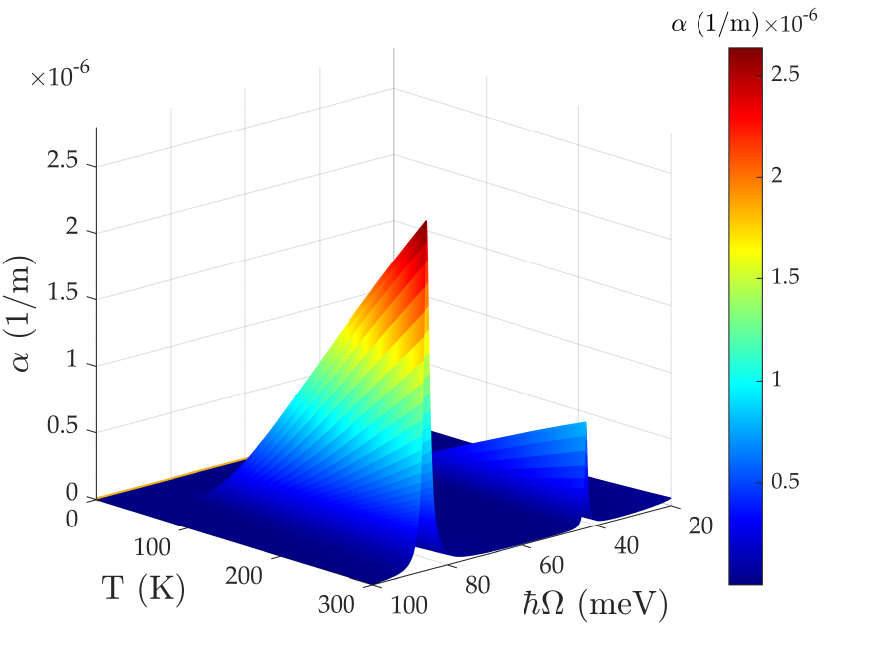}}
\subfigure[][\label{7b}]
  {\includegraphics[width=0.47\linewidth]{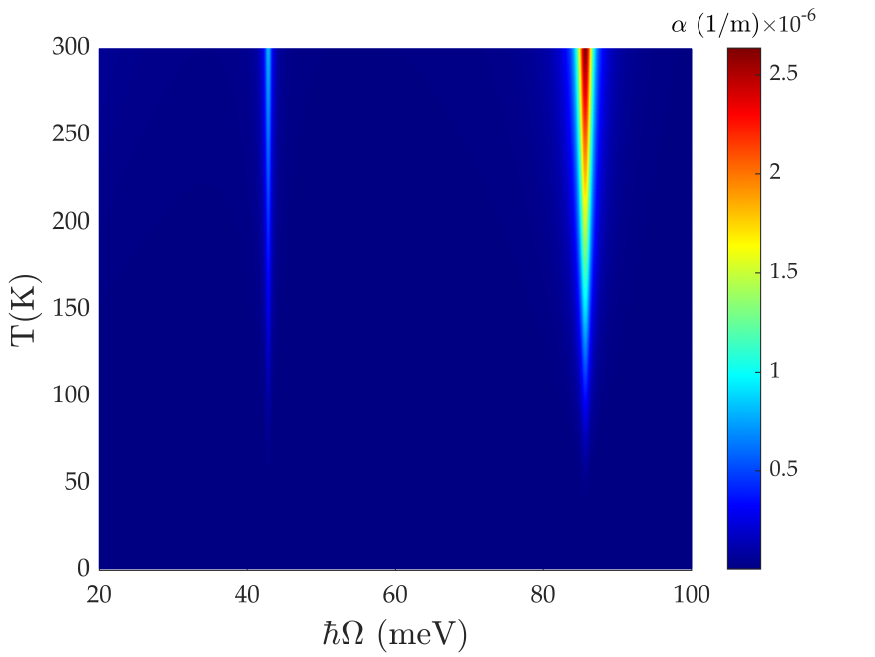}}
\caption{(Color online) TAC as a function of the temperature and the photon energy with the side view (left) and the top view (right). Here, $B=16$T, $\omega_z=0.1\omega_0$, and $E_0=2 \times 10^5 \rm{V/m}$.}
\label{fig7}
\end{figure}

\begin{figure}[!htb]
\centering
\subfigure[][\label{8a}]
  {\includegraphics[width=0.47\linewidth]{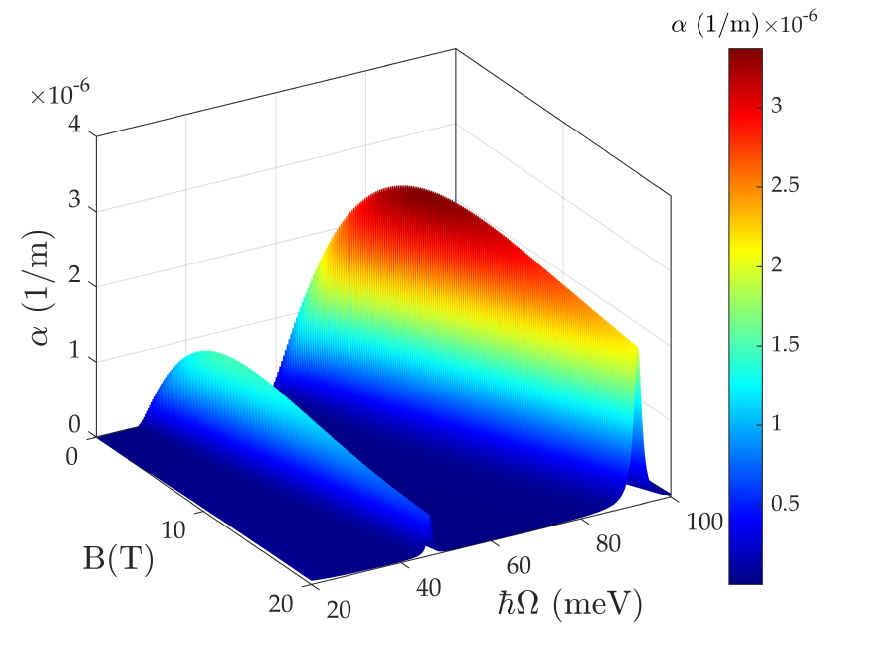}}
\subfigure[][ \label{8b}]
  {\includegraphics[width=0.47\linewidth]{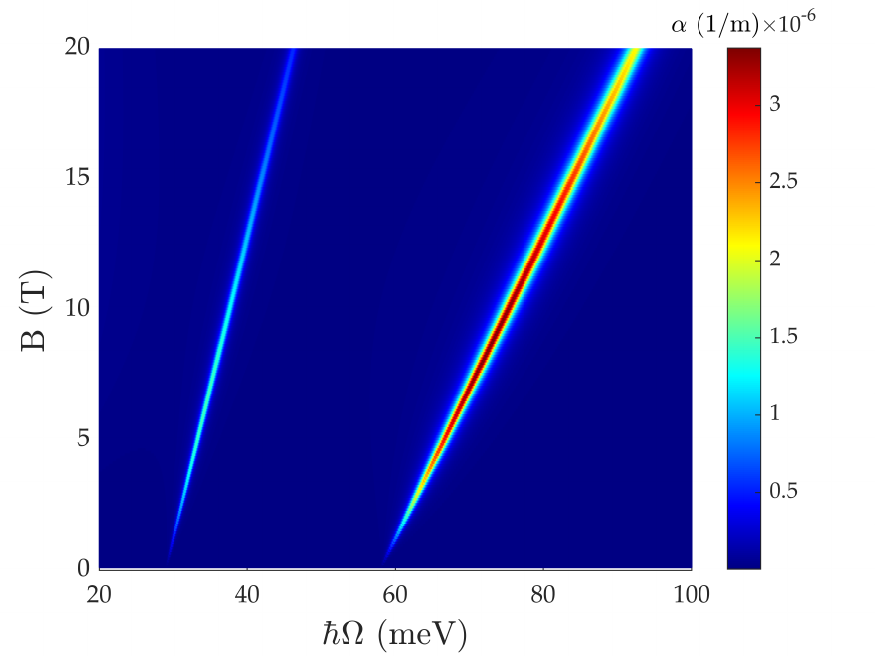}}
\caption{(Color online) TAC as a function of the magnetic field and the photon energy with the side view (left) and the top view (right). Here, $T=200$K, $\omega_z=0.1\omega_0$, and $E_0=2 \times 10^5 \rm{V/m}$.}
\label{fig8}
\end{figure}

\begin{figure}[!htb]
\centering
\subfigure[][\label{9a}]
  {\includegraphics[width=0.47\linewidth]{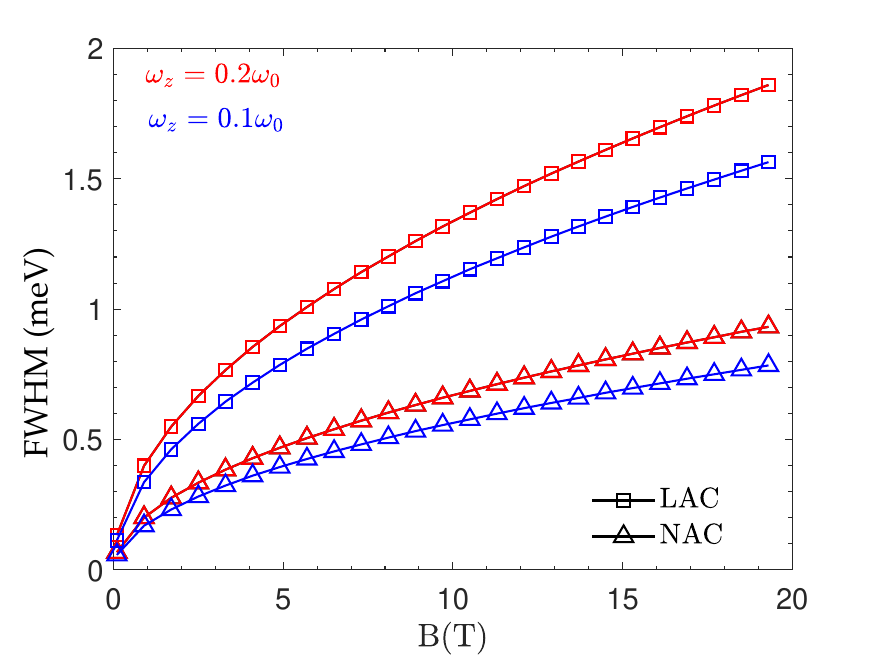}}
\subfigure[][ \label{9b}]
  {\includegraphics[width=0.47\linewidth]{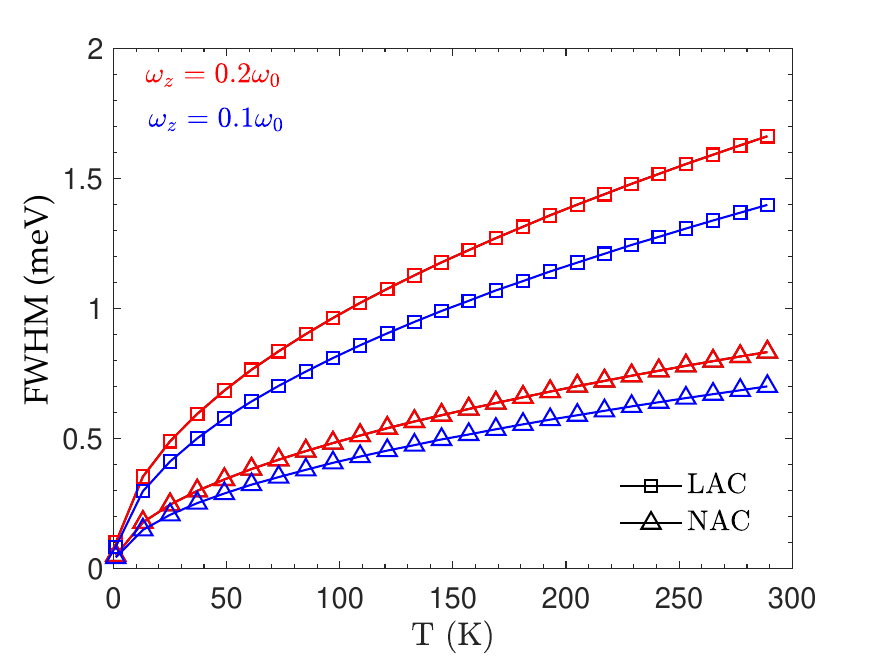}}
  \subfigure[][ \label{9c}]
  {\includegraphics[width=0.47\linewidth]{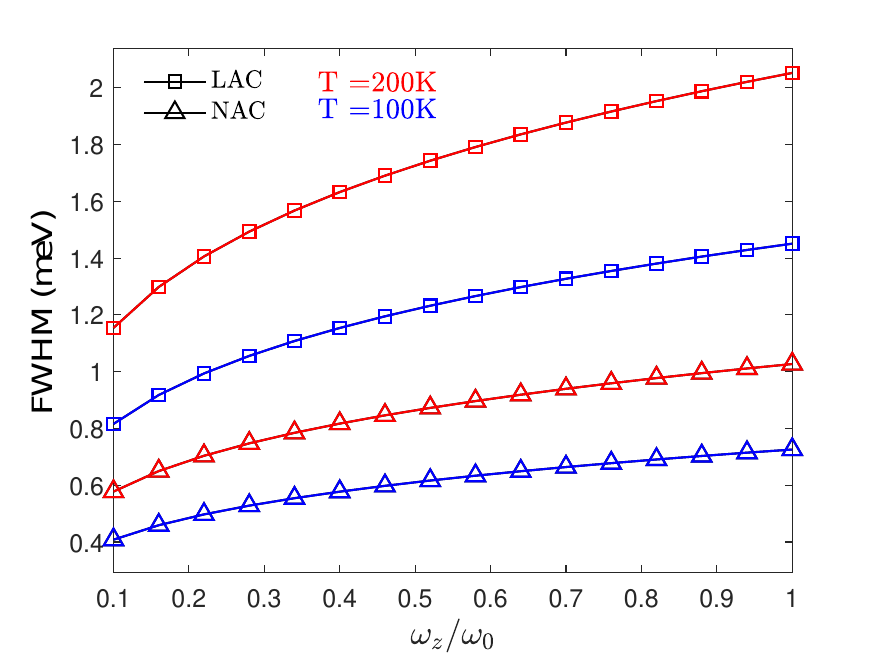}}
  \subfigure[][ \label{9d}]
  {\includegraphics[width=0.47\linewidth]{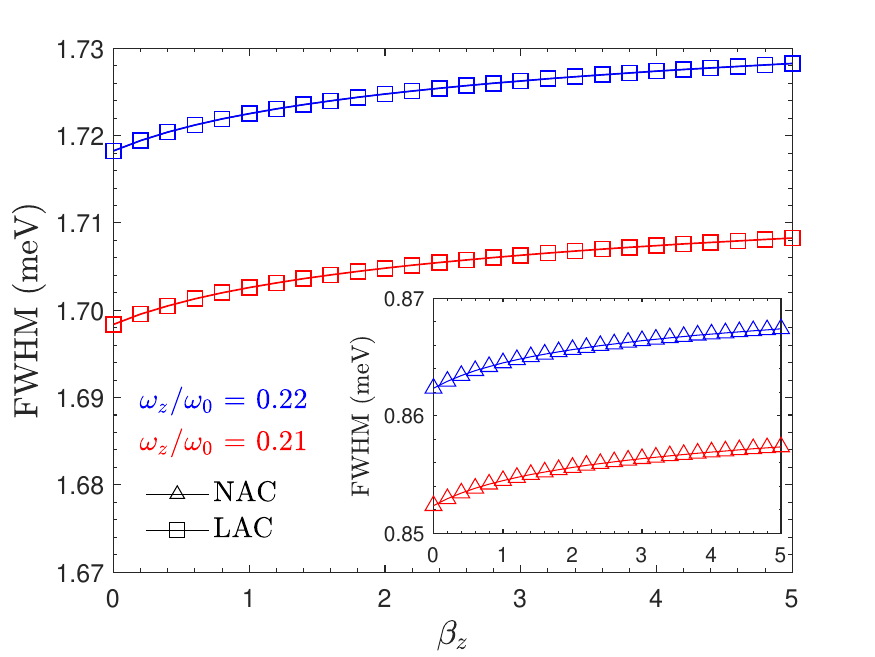}}
\caption{(Color online) FWHM of LAC and NAC as a function of the magnetic field, the ratio ${{{\omega _z}} \mathord{\left/
 {\vphantom {{{\omega _z}} {{\omega _0}}}} \right.
 \kern-\nulldelimiterspace} {{\omega _0}}}$, the temperature and $\beta_z$.}
\label{fig9}
\end{figure}

\begin{table}[!htb]
\centering
\begin{tabular}{|l|cccc|}
\hline
\multicolumn{1}{|c|}{\multirow{3}{*}{\textbf{}}} & \multicolumn{4}{c|}{\begin{tabular}[c]{@{}c@{}}Figure 9(a)\\ $\rm{FWHM} = {\kappa _a}\sqrt{B}+{\theta}_{a} (\rm{meV,T})$\end{tabular}}                               \\ \cline{2-5} 
\multicolumn{1}{|c|}{}                           & \multicolumn{2}{c|}{${\kappa_a}$ ($\rm{meV}/\sqrt{T}$)}                                     & \multicolumn{2}{c|}{$\theta_a$ ($\times 10^{-4}$meV)}                  \\ \cline{2-5} 
\multicolumn{1}{|c|}{}                           & \multicolumn{1}{c|}{$\omega_z=0.1\omega_0$}  & \multicolumn{1}{c|}{$\omega_z=0.2\omega_0$}  & \multicolumn{1}{c|}{$\omega_z=0.1\omega_0$}  & $\omega_z=0.2\omega_0$  \\ \hline
\multicolumn{1}{|c|}{LAC}                        & \multicolumn{1}{c|}{0.3559}                  & \multicolumn{1}{c|}{0.4236}                  & \multicolumn{1}{c|}{4.5346}                  & - 9.5443                \\ \hline
\multicolumn{1}{|c|}{NAC}                        & \multicolumn{1}{c|}{0.1783}                  & \multicolumn{1}{c|}{0.2122}                  & \multicolumn{1}{c|}{6.1038}                  & - 9.3563                \\ \hline
\multirow{3}{*}{}                                & \multicolumn{4}{c|}{\begin{tabular}[c]{@{}c@{}}Figure 9(b)\\ $\rm{FWHM} = {\kappa_b}\sqrt{T}+{\theta_b} (\rm{meV,K})$\end{tabular}}                                  \\ \cline{2-5} 
                                                 & \multicolumn{2}{c|}{$\kappa_b$ ($\rm{meV}/\sqrt{K}$)}                                       & \multicolumn{2}{c|}{$\theta_b$ ($\times 10^{-4}$meV)}                  \\ \cline{2-5} 
                                                 & \multicolumn{1}{c|}{$\omega_z=0.1\omega_0$}  & \multicolumn{1}{c|}{$\omega_z=0.2\omega_0$}  & \multicolumn{1}{c|}{$\omega_z=0.1\omega_0$}  & $\omega_z=0.2\omega_0$  \\ \hline
\multicolumn{1}{|c|}{LAC}                        & \multicolumn{1}{c|}{0.0833}                  & \multicolumn{1}{c|}{0.0977}                  & \multicolumn{1}{c|}{8.125}                   & 13                      \\ \hline
\multicolumn{1}{|c|}{NAC}                        & \multicolumn{1}{c|}{0.0411}                  & \multicolumn{1}{c|}{0.0489}                  & \multicolumn{1}{c|}{8.75}                    & 11                      \\ \hline
\multirow{3}{*}{}                                & \multicolumn{4}{c|}{\begin{tabular}[c]{@{}c@{}}Figure 9(c)\\ $\rm{FWHM} = {\kappa_c}\sqrt{\dfrac{\omega_z}{\omega_0}}+{\theta_c} (\rm{meV})$\end{tabular}}           \\ \cline{2-5} 
                                                 & \multicolumn{2}{c|}{$\kappa_c$ (meV)}                                                       & \multicolumn{2}{c|}{$\theta_c$ (meV)}                                  \\ \cline{2-5} 
                                                 & \multicolumn{1}{c|}{$\rm{T}=100\rm{K}$}      & \multicolumn{1}{c|}{$\rm{T}=200\rm{K}$}      & \multicolumn{1}{c|}{$\rm{T}=100\rm{K}$}      & $\rm{T}=200\rm{K}$      \\ \hline
\multicolumn{1}{|c|}{LAC}                        & \multicolumn{1}{c|}{0.9287}                  & \multicolumn{1}{c|}{1.3136}                  & \multicolumn{1}{c|}{0.5224}                  & 0.7391                  \\ \hline
\multicolumn{1}{|c|}{NAC}                        & \multicolumn{1}{c|}{0.4645}                  & \multicolumn{1}{c|}{0.6572}                  & \multicolumn{1}{c|}{0.2613}                  & 0.3699                  \\ \hline
\multirow{3}{*}{}                                & \multicolumn{4}{c|}{\begin{tabular}[c]{@{}c@{}}Figure 9(d)\\ $\rm{FWHM} = {\kappa_d}\sqrt{\beta_z}+{\theta_d} (\rm{meV})$\end{tabular}}                              \\ \cline{2-5} 
                                                 & \multicolumn{2}{c|}{$\kappa_d$ (meV)}                                                       & \multicolumn{2}{c|}{$\theta_d$ (meV)}                                  \\ \cline{2-5} 
                                                 & \multicolumn{1}{c|}{$\omega_z=0.21\omega_0$} & \multicolumn{1}{c|}{$\omega_z=0.22\omega_0$} & \multicolumn{1}{c|}{$\omega_z=0.21\omega_0$} & $\omega_z=0.22\omega_0$ \\ \hline
LAC                                              & \multicolumn{1}{c|}{0.0044}                  & \multicolumn{1}{c|}{0.0045}                  & \multicolumn{1}{c|}{1.6984}                  & 1.7183                  \\ \hline
NAC                                              & \multicolumn{1}{c|}{0.0022}                  & \multicolumn{1}{c|}{0.0023}                  & \multicolumn{1}{c|}{0.8524}                  & 0.8524                  \\ \hline
\end{tabular}
\caption{Law of dependence of FWHM on the magnetic field B, temperature T, confinement frequency $\omega_z$ and $\beta_z$ of ISPPSISQW.}
\label{quyluat}
\end{table}
 Finally, in Fig. \ref{fig9}, we provide some detailed calculations on the dependence of the FWHM of the one (LAC) and two-photon absorption (NAC) peaks on several external parameters such as magnetic field, temperature, and confinement frequency of ISPPSISQW. The dependence of FWHM on the above quantities follows the square root law given in Table \ref{quyluat}. The dependence of FWHM on the quantity ${\rm{X}}\left( {{\rm{B}}{\rm{,T}}{\rm{,}}\dfrac{{{\omega _z}}}{{{\omega _0}}}{\rm{,}}{\beta_z}} \right)$ has the form of a square root ${\rm{FWHM}} = \kappa \sqrt {\rm{X}}  + \theta $. This is also consistent with previous studies on the nonlinear absorption of SEMW in Quantum Wells with different potential profiles \cite{tung,tung2,lvtung} and in Graphene \cite{tuan2023, phuc}. In contrast to the case without an external magnetic field (see Fig. \ref{4a}), the FWHM depends strongly and increases monotonically with the confinement frequency according to the square root law shown in Fig. \ref{9c}. This can be explained by the influence of the external magnetic field causing the appearance of a specific harmonic confinement potential, increasing the probability of electron-phonon scattering, and leading to a sharp increase of FWHM with the confinement frequency of ISPPSISQW. To our knowledge, these results are new, built on the basis of the quantum kinetic equation method (quantum theory). Qualitatively, the rule about the dependence of FWHM on the square root magnetic field is true for all two-dimensional systems, especially this rule for two-dimensional Graphene is in good agreement with theoretical calculations \cite{tuan2023,phuc,hoi1} and experimental observations \cite{ji,orlita}. Quantitatively, the FWHM calculation results in two-dimensional ISPPSIS Quantum Wells are different from two-dimensional Graphene at the parameter values $\kappa$ and $\theta$ in the law depending on the magnetic field of the form ${\rm{FWHM}} = \kappa \sqrt {\rm{B}}  + \theta$. This is explained by the difference in the nature of the electronic confinement potential and different material parameters even though ISPPSIS Quantum Wells and Graphene are two-dimensional systems. In terms of detailed quantification, the estimates we present in Table 1 for $\kappa$ and $\theta$ can be used as prediction criteria in future experimental observations.

\section{Conclusions}

In this paper, we analytically investigated the nonlinear absorption of an SEMW by electrons confined in ISPPSISQW by using the quantum kinetic equation. We have obtained the general TAC of SEMW in ISPPSISQW (electron-optical phonon scattering) from the low-temperature to the high-temperature domain in both the absence and presence of an external magnetic field. The general analytical expression obtained depends on the parameters of strong electromagnetic waves: the intensity of SEMW $E_{0}$, photon energy $\hbar {\Omega}$, external parameters such as temperature T and the magnetic field B, and characteristic parameters of the Quantum Wells such as the confinement frequency $\omega_z$ and $\beta_z$. In the case of magnetic field absence, the numerical results show that as the temperature of the system increases, the TAC increases. This is similar to the case of bulk semiconductors \cite{eps} and low-dimensional electronic systems previously investigated \cite{van, tung, hien, bau}. However, if the influence of SEMW is increased, namely increasing both the strength and frequency of SEMW, the TAC also increases significantly. An important point in this work is that the electron-phonon resonance condition helps to precisely locate the peaks of the electron's electromagnetic wave absorption spectrum. Besides, the FWHM is almost independent of the confinement frequency and increases monotonically with temperature. The obtained results can be used as important evaluation criteria in future studies and experimental observations. 

In the presence of an external magnetic field, the electron absorption spectrum is obtained using magneto-phonon resonance conditions. The results show that the position of the resonance peaks does not depend on temperature. However, temperature strongly affects the value of the resonance peaks. When the temperature increases, the resonance peak value also increases. The one-photon absorption peak value is 2.2 times higher than the two-photon absorption peak value. The position of the resonance peaks is also shifted to the right when increasing the external magnetic field. By detailed numerical calculations, we also obtain the dependence of FWHM on the quantity ${\rm{X}}\left( {{\rm{B}}{\rm{, T}}{\rm{,}}\dfrac{{{\omega _z}}}{{{\omega _0}}}{\rm{,}}{\beta_z}} \right)$  with a square root law. These results are consistent with previous studies on Quantum Wells with different potential types. The detailed calculations can serve as important suggestions for future experimental observations. 

Our approach - the quantum kinetic equation method used in this paper can be used to study the quantum theory for physical properties of many-body systems such as semiconductor systems, advanced material systems, two-dimensional systems, and one-dimensional systems, … which includes two-dimensional Graphene and two-dimensional ISPPSIS Quantum Wells. However, the resulting analytical equations here are not valid for general patterns such as semiconductor systems, advanced material systems, two-dimensional systems, and one-dimensional systems, … This can be explained by the difference in wave function and energy spectrum of electrons and the interactions between electrons and other particles in the above different systems, leading to different analytical expressions of MPNAC.
\section*{Acknowledgement}
This research is financial by Vietnam National University, Hanoi - Grant number QG. 23.06.

\bibliography{refs}
\newpage
\end{document}